\documentclass{article}

\usepackage{arxiv}

\usepackage[dvipsnames]{xcolor}
\usepackage[edges]{forest}
\usepackage{tikz}
\usepackage[hidelinks]{hyperref}       
\usepackage{subcaption}
\usepackage{caption}
\usepackage{amsmath}
\usepackage{amsfonts}
\usepackage{amssymb}
\usepackage[T1]{fontenc}
\usepackage{microtype}
\usepackage{booktabs}
\usepackage{nicefrac}
\usetikzlibrary{decorations.pathmorphing, shapes, positioning, arrows, arrows.meta}

\tikzset{snake it/.style={decorate, decoration=snake}}

\author{{Robert J. Hayek, Joaquin Chung, Rajkumar Kettimuthu\thanks{Corresponding authors: Robert Hayek (email: rhayek@anl.gov) and Rajkumar Kettimuthu (email:kettimut@anl.gov,
Project PI).}}\\
	Data Science and Learning\\
	Argonne National Laboratory\\
	Lemont, IL, USA 60439 \\
	\texttt{\{rhayek, chungmiranda, kettimut\}@anl.gov} \\
}

\hypersetup{
pdftitle={A Review of Software for Designing and Operating Quantum Networks},
pdfsubject={quant-ph},
pdfauthor={Robert J. Hayek, Joaquin Chung, Rajkumar Kettimuthu},
pdfkeywords={Quantum Networks, Protocols, Network Design, Software Design, Simulation},
}

\begin{document}
\title{A Review of Software for Designing and Operating Quantum Networks}

\maketitle

\begin{abstract}
Quantum networks development is crucial to realizing a production-grade network that can support distributed sensing, secure communication, and utility-scale quantum computation. 
However, the transition from laboratory demonstration to deployable
networks requires software implementations of 
architectures and protocols tailored to the unique constraints of quantum systems. 
This paper reviews the current state of software implementations for quantum networks, organized around a three-plane abstraction of infrastructure, logical, and control/service planes. 
We cover software for both designing quantum network protocols (e.g., SeQUeNCe, QuISP, and NetSquid) and operating testbeds, with a focus on essential control/service plane functions such as entanglement, topology, and resource management, in a proposed taxonomy. 
Our review highlights a persistent gap between theoretical architecture and protocol proposals and their realization in simulators or testbeds, particularly in dynamic topology and network management. We conclude by outlining open challenges and proposing a roadmap for developing scalable software architectures to enable hybrid, large-scale quantum networks.
\end{abstract}


\section{Introduction} \label{sec:intro}
Quantum networks~\cite{kimble2008quantum,wehner2018quantum} promise more secure communication protocols~\cite{deutsch_quantum_1996}, precise distributed sensing~\cite{zhang2021distributed}, and better computational capabilities~\cite{cuomo2020towards} by connecting multiple quantum resources. Despite successful demonstration of quantum communication at laboratory and metropolitan scales~\cite{LIU2025100551}, today's quantum networks remain in their infancy. Building reliable, multiuser networks over arbitrary distances will require integrating diverse quantum and classical components~\cite{van2022quantum}. 

The research community has therefore pursued two complementary approaches: (1) simulators that model the behavior of quantum networks and (2) small-scale experimental testbeds. Both approaches depend on software tailored to the needs of quantum networks. The main function of a quantum network is to provide entanglement distribution services to users and applications across a wide range of geographical scales, spanning from interchip to data center to wide-area networks. Achieving this requires coordination among devices, such as quantum memories, quantum frequency converters, Bell state analyzers, and entangled photon sources, all of which operate under tight synchronization constraints. This is akin to current trends in classical networking, where monolithic devices have been disaggregated into individual functions that are composed into services via software orchestration~\cite{nfv-sdn-survey2019}.

Early proposals for a quantum internet~\cite{dahlberg_link_2019,alshowkan2021,li_building_2021,li_survey_2024} attempted to map directly to the TCP/IP stack~\cite{forouzan2002tcp}, but Illiano et al.~\cite{qi-stack-survey} and others~\cite{pirker_resource-centric_2025} have emphasized the limitations of direct translation given the requirements of entanglement distribution. In this paper we review software for quantum networks using the three-plane abstraction (infrastructure, logical, and control/service) to organize state-of-the-art developments without prescribing yet another protocol stack. Our primary contribution is to classify the essential functions that a quantum network must execute to provide entanglement distribution reliably, and to survey how these functions are realized in simulations and, where possible, in experimental testbeds.

The rest of the paper is organized as follows. Section~\ref{sec:background} provides background on quantum networks and useful analogies to modern, classical networking systems. Sections \ref{sec:sw-design} and \ref{sec:sw-ops} review the existing software for both designing and operating quantum networks, respectively.  Section~\ref{sec:discussion} discusses open research areas.
\section{Background}\label{sec:background}
This section provides background on quantum networking basic functions, their synchronization constraints, and the plane abstraction adopted from classical networks.

\subsection{Quantum Networking Basic Functions}
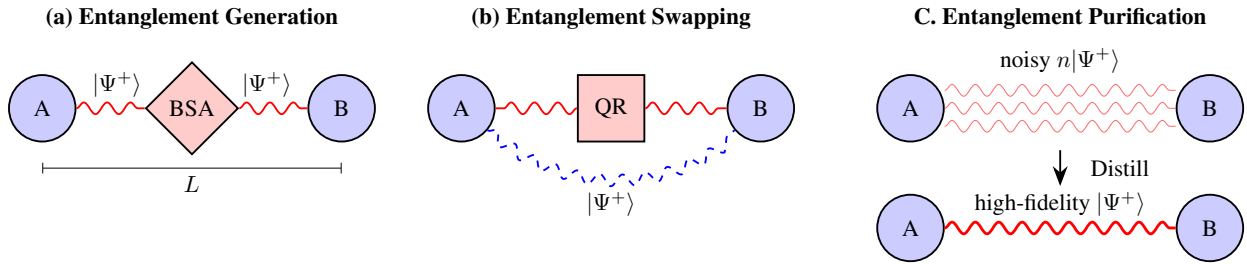
\begin{figure}[htbp]
    \resizebox{\columnwidth}{!}{%
    \centering
    \begin{tikzpicture}[
            node/.style={circle, draw, thick, minimum size=1cm, fill=blue!20},
            repeater/.style={draw, thick, minimum size=1cm, fill=red!20},
            bsm/.style={diamond,draw, thick, minimum size=1cm, fill=red!20},
            entangle/.style={thick, red, snake it},
            classical/.style={thick, blue, dashed},
            arrow/.style={-{Stealth[length=2mm]}},
            scale=0.9
        ]

        \node at (2.5, 4.5) {\textbf{(a) Entanglement Generation}};
        \node[node] (A1) at (0, 3) {A};
        \node[bsm] (R1) at (2.5, 3) {BSA};
        \node[node] (B1) at (5, 3) {B};

        \draw[entangle] (A1) -- (R1);
        \draw[entangle] (R1) -- (B1);
        \node at (1.25, 3.4) {$|\Psi^+\rangle$};
        \node at (3.75, 3.4) {$|\Psi^+\rangle$};

        \draw[|-|, thin] (0, 2) -- (5, 2) node[below, midway]{$L$};
        
        \node at (9.5, 4.5) {\textbf{(b) Entanglement Swapping}};
        \node[node] (A2) at (7, 3) {A};
        \node[repeater] (R2) at (9.5, 3) {QR};
        \node[node] (B2) at (12, 3) {B};

        \draw[entangle, dashed, blue] (A2) to[out=-40, in=-140] (B2);
        \draw[entangle] (R2) -- (A2);
        \draw[entangle] (R2) -- (B2);
        \node at (9.5, 1.4) {$|\Psi^+\rangle$};

        \node at (17, 4.5) {\textbf{C. Entanglement Purification}};
        \node[node] (A3) at (14.5, 3) {A};
        \node[node] (B3) at (19.5, 3) {B};

        \draw[entangle, thin, opacity=0.6] (A3) -- (B3);
        \draw[entangle, thin, opacity=0.6] ([yshift=3mm]A3.east) -- ([yshift=3mm]B3.west);
        \draw[entangle, thin, opacity=0.6] ([yshift=-3mm]A3.east) -- ([yshift=-3mm]B3.west);
        \node at (17, 3.8) {noisy $n|\Psi^+\rangle$};

        \draw[-{Stealth[length=3mm]}, thick] (17, 2.3) -- (17, 1.7);
        \node at (18, 2) {Distill};

        \node[node] (A4) at (14.5, 1) {A};
        \node[node] (B4) at (19.5, 1) {B};
        \draw[entangle, very thick] (A4) -- (B4);
        \node at (17, 1.4) {high-fidelity $|\Psi^+\rangle$};

        \end{tikzpicture}%
    }
    \caption{Basic entanglement functions: (a) entanglement generation using the meet-in-the-middle (MIM) protocol, (b) quantum-repeater-enabled entanglement swapping, and (c) entanglement purification. BSA: Bell state analyzer, QR: quantum repeater.}
    \label{fig:entanglement_funcs}
\end{figure}

To distribute entanglement between users, quantum networks rely on three basic functions: entanglement generation,  swapping, and purification. Entanglement generation protocols~\cite{barrett_efficient_2005} create quantum entanglement between two qubits of adjacent nodes in a quantum network.  Two main protocols for distributing entanglement between adjacent nodes have been proposed: meet-in-the-middle (MIM) and midpoint source (MS)~\cite{jones_design_2016}.  In MIM (see Fig.~\ref{fig:entanglement_funcs}(a)), two nodes are separated by a distance $L$, with a Bell state analyzer (BSA) placed at the midpoint ($L/2$) of the nodes. Each node synchronously generates a photon-qubit Bell pair, where one photon from each pair is transmitted toward the central BSA. The probability of successful entanglement delivery in MIM scales as $p_{\text{ent}} \propto p^2_{\text{optical}}$, where $p_{\text{optical}}$ is the transmission probability over distance $L/2$, as two independent photons must successfully traverse their channel. In the MS protocol (not shown in Fig.~\ref{fig:entanglement_funcs}(a)), two adjacent nodes are separated by a distance $L$, but the BSA is replaced by an entangled photon source (EPS) located at the midpoint. As the name suggests, an entangled photon source creates pairs of entangled photons via nonlinear optical processes such as spontaneous parametric down conversion (SPDC)~\cite{scholz2009narrowband}.

In order to enable long-distance entanglement generation between nodes in a quantum network, quantum repeaters (see Fig.~\ref{fig:entanglement_funcs}(b)) are required to enable long-distance quantum communications. The reason is that  the entanglement generation rate decays with distance~\cite{takeoka_fundamental_2014}. More specifically, the entanglement generation rate that can be achieved without a repeater is defined by the \textit{PLOB} bound, namely, $R=-\log_2(1-T)$, where $T$ is the transmission of the link~\cite{pirandola_fundamental_2017}.
This operation, which relies on protocols for Bell and multipartite state~\cite{ji_entanglement_2022} measurements, generates entanglement between distant pairs without direct interaction. 
The generation of entanglement occurs over noisy channels that cause photon loss and degradation of the fidelity of entangled pairs. 
In order to address this situation, entanglement purification protocols (see Fig.~\ref{fig:entanglement_funcs}(c)) are implemented to distill higher-fidelity pairs from an ensemble of lower-fidelity pairs (for initial fidelity $\geq0.5$ for mixed and Werner state pairs)~\cite{deutsch_quantum_1996,bennett_purification_1996,rozpedek_optimizing_2018}.

\subsection{Synchronized Operations}
The previously described functions require  precise, synchronized operations to succeed.
Photonic entanglement generation protocols, for example, depend on Hong--Ou--Mandel (HOM) interference, where a HOM interferometer is used to measure the amount of indistinguishability between two input photons. 
The measurement of this ``indistinguishability'' is the visibility $V$, where $0 \leq V \leq 1$. 
The visibility of the interference measurement can be affected by polarization, spectral modes, temporal modes, arrival time, and transverse spatial mode. 
Of importance for timing and synchronization is the arrival time of photons. 

Many components of quantum networks impose tight timing constraints. 
For instance, wave packet overlap is when two incident photons of a beam splitter differ in their arrival time. 
If the arrival times of both incident photons of a HOM interferometer differ, then the coincidence detection probability will significantly decrease.
In the case of single-photon sources, if emitted photons are spectrally pure, then the uncertainty in emission time becomes large because the emission time of a spectrally pure photon is a Poisson distribution with a standard deviation of $\sim10\mu{s}$, expressed as jitter. 
The detection efficiency of single-photon detectors depends on their recovery time, dark count rate, and timing jitter, all of which contribute to timing constraints of entanglement generation~\cite{nagayama}.


\subsection{Plane Abstraction and Network Functions} \label{sec:planes}

\begin{figure}
    \centering
    \includegraphics[width=0.6\textwidth]{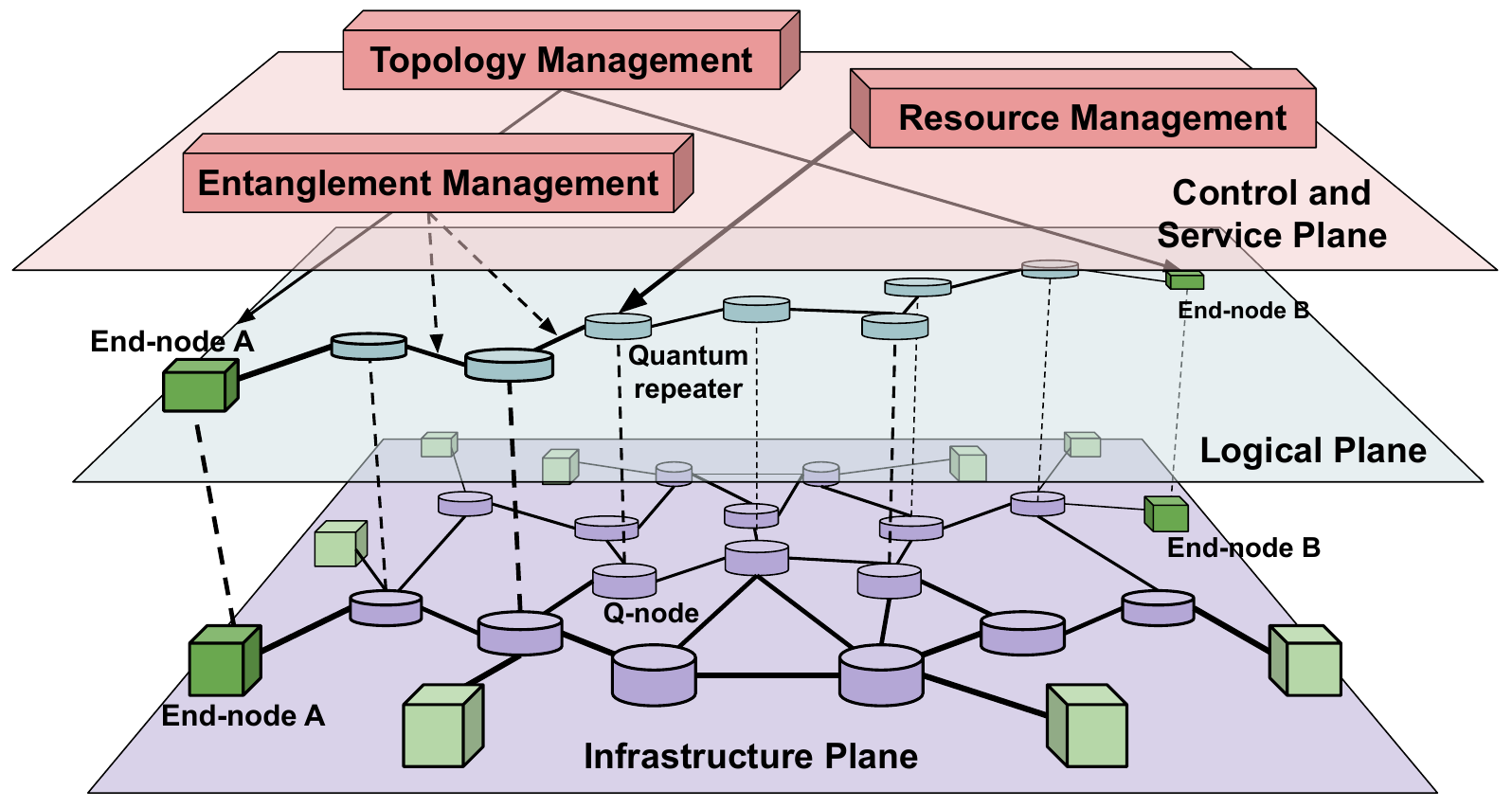}
    \caption{Architecture of a quantum network using a plane abstraction. The control and service plane include network functions such as entanglement, topology, and resource management (see Section~\ref{sec:sw-ops}). Adapted under the terms of the CC-BY license~\cite{chung_interqnet_2025}. Copyright 2025, The Author(s).}
    \label{fig:plane-qnet}
\end{figure}

Figure~\ref{fig:plane-qnet} illustrates a quantum network organized using a three-plane abstraction described in~\cite{chung_interqnet_2025}. The infrastructure plane at the bottom represents all physical devices and fiber connections. The logical plane in the middle captures a quantum-repeater-chain abstraction once a network orchestrator has identified a suitable path between end nodes (e.g., A and B in dark green). 
This logical plane could also abstract the artificial, entanglement-based topology described in~\cite{mazza-intra-qlan-2025}.
The control and service plane at the top encompasses the network functions and user-facing services. 
Similar to software-define networking (SDN)~\cite{peterson2021sdn}, the control and service plane of quantum networks will interact with users/applications via Northbound APIs and with the logical or infrastructure plane via Southbound APIs.
Examples of quantum network functions are entanglement, topology, and resource managements. Section~\ref{sec:sw-ops} describes these functions (and their subfunctions) in detail.


\section{Software for Designing Quantum Networks} \label{sec:sw-design}
\begin{figure}[tbhp]
    \centering
    \begin{subfigure}[b]{0.3\textwidth}
        \centering
        \includegraphics[width=\textwidth]{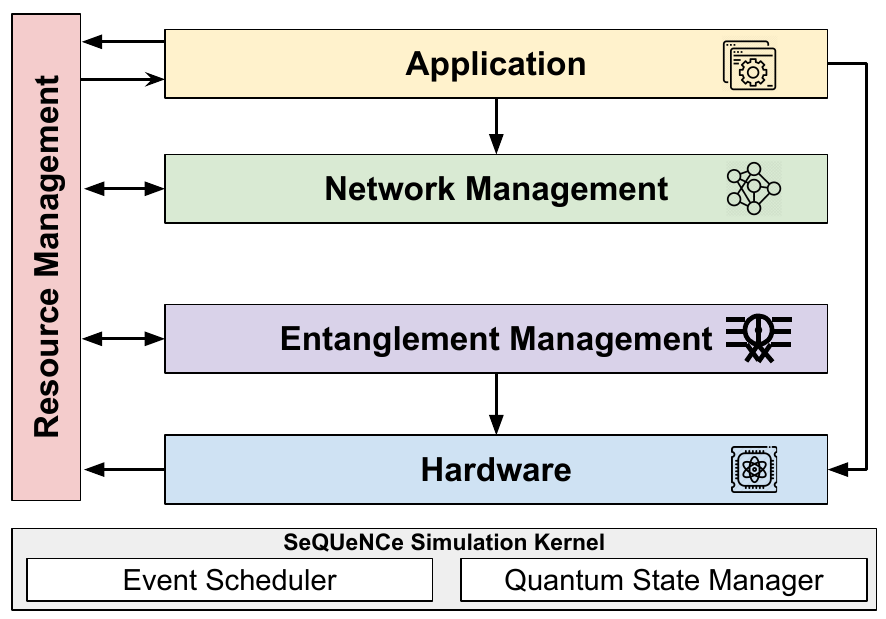}
        \caption{SeQUeNCe. Reproduced under the terms of the CC-BY license~\cite{wu_sequence_2021}. Copyright 2021, The Author(s).}
        \label{fig:sequence_arch}
    \end{subfigure}
    \hfill
    \begin{subfigure}[b]{0.3\textwidth}
        \centering
        \includegraphics[width=\textwidth]{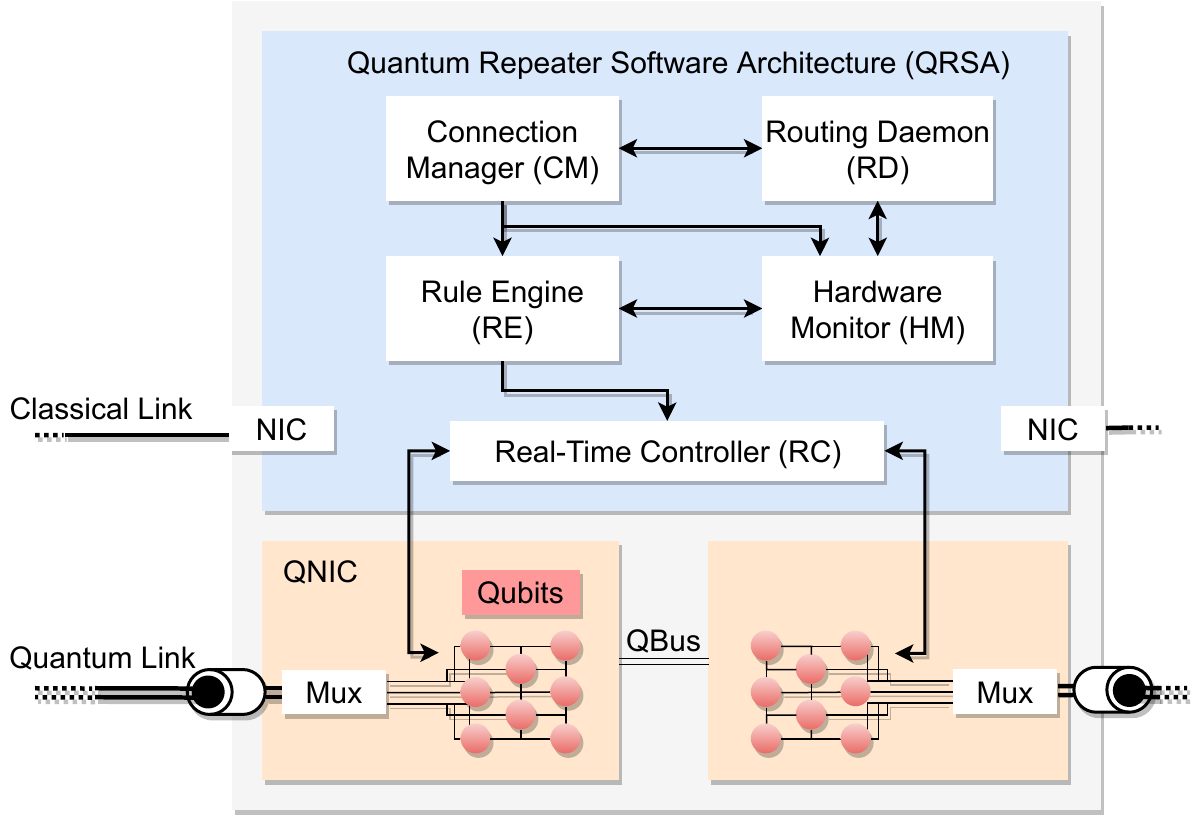}
        \caption{QuISP. Reproduced under the terms of the CC-BY license~\cite{satoh_quisp_2021}. Copyright 2021, The Author(s).}
        \label{fig:quisp_arch}
    \end{subfigure}
    \hfill
    \begin{subfigure}[b]{0.3\textwidth}
        \centering
        \includegraphics[width=\textwidth]{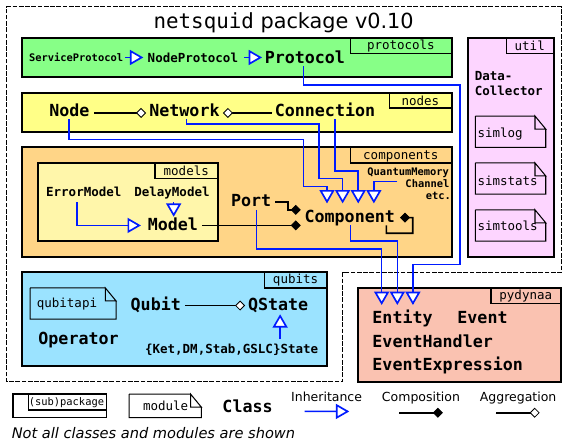}
        \caption{NetSquid. Reproduced under the terms of the CC-BY license~\cite{coopmans_netsquid_2021} Copyright 2021, The Author(s).}
        \label{fig:netsquid_arch}
    \end{subfigure}

    \caption{Architectures for the (a) SeQUeNCe, (b) QuISP, and (c) NetSquid quantum network simulators.}
    \label{fig:arch_grid}
\end{figure}

Because of the technical immaturity of quantum networking hardware, testing new quantum network software implementations is challenging. Thus, simulators are crucial for designing and verifying the correctness and scalability of new quantum network software proposals. Many simulators for quantum networks have been developed recently, each with a specific goal. Here we summarize popular simulators and consolidate the system architecture diagrams for major simulators in Fig.~\ref{fig:arch_grid}. A quantitative performance benchmark of these simulators is out of the scope of this review. Nevertheless, we extend Table 1 in Ref.~\cite{cross-val2025} to qualitatively compare major simulators used in academic research (see Table~\ref{tab:sim-comparison}).

SeQUeNCe~\cite{wu_sequence_2021} is an open-source discrete-event simulator written in Python that simulates the evolution of quantum states encoded in photons as they traverse a quantum network with arbitrary topologies. SeQUeNCe implements a software framework consisting of modular components, including the resource, network, and entanglement managers (see Fig.~\ref{fig:sequence_arch}), to create a  broad quantum network architecture. The simulator allows new protocols to be directly plugged-in to the software framework for experimentation. It also supports various built-in entanglement generation, purification, and swapping protocols, along with managers for each, while also implementing multiple formalisms for the quantum state. SeQUeNCe strives to maintain a balance between accuracy of simulations, scale of the simulated networks, and usability.

QuISP~\cite{satoh_quisp_2022} is an open-source simulator for quantum networks written in OMNeT++, a C++ library for discrete event simulation. It implements a full network stack, with photonic entanglement between peers. Its purpose is to help researchers understand the behavior of quantum networks at large scales. Shown in Fig.~\ref{fig:quisp_arch}, the simulator implements a ``quantum repeater software architecture'' (QRSA),  which is a modular network architecture consisting of a connection manager, hardware monitor, rule engine, real-time controller, and routing daemon.

NetSquid is a closed-source (but open-API) discrete-event simulator providing capabilities to simulate all aspects of quantum computing and a quantum network stack~\cite{coopmans_netsquid_2021}. Its main focus is to enable research in large-scale quantum network dynamics. The simulator utilizes a modular design, with components for protocols and the physical models using the pydynaa discrete-event simulator package. Figure~\ref{fig:netsquid_arch} presents NetSquid's software architecture.

QuantumSavory is a quantum network simulator written in Julia to study quantum networks, with a special focus on physical layer dynamics. It offers both a stabilizer and state vector quantum state formalisms as simulation backends; both of these engines are provided by external libraries. The simulator includes noise and error models to allow for the evolution of quantum states affected by noise and decoherence. Additionally, the QuantumSavory developers have implemented a library of built-in protocols, states, and quantum circuits. 

\begin{table}[]
\centering
\caption{Comparison of quantum network simulator design and implementation choices.}
\label{tab:sim-comparison}
\resizebox{\textwidth}{!}{%
\begin{tabular}{|l|l|c|l|c|l|l|}
\hline
 &
  \multicolumn{1}{c|}{\textbf{Goal}} &
  \textbf{Open Source} &
  \multicolumn{1}{c|}{\textbf{Platform}} &
  \textbf{\begin{tabular}[c]{@{}c@{}}Quantum\\ Network\\ Gen.~\cite{muralidharan2016optimal}\end{tabular}} &
  \multicolumn{1}{c|}{\textbf{Software Architecture}} &
  \multicolumn{1}{c|}{\textbf{Error Models}} \\ \hline
\textbf{QuISP} &
  \begin{tabular}[c]{@{}l@{}}To simulate large-scale quantum \\ network and internetworks.\end{tabular} &
  Yes &
  \begin{tabular}[c]{@{}l@{}}OMNeT++ \\ and C++\end{tabular} &
  1G &
  \begin{tabular}[c]{@{}l@{}}Each node in the network implements\\ QRSA.\end{tabular} &
  Support for qubit error models. \\ \hline
\textbf{SeQUeNCe} &
  \begin{tabular}[c]{@{}l@{}}To accurately simulate photonic \\ quantum networks, while providing \\ high levels of customizability.\end{tabular} &
  Yes &
  Python 3.10+ &
  1G &
  \begin{tabular}[c]{@{}l@{}}Simulation kernel plus applications,\\ hardware, entanglement, network,\\ and resource management modules.\end{tabular} &
  \begin{tabular}[c]{@{}l@{}}Errors are encoded as \\ variable hardware parameters.\end{tabular} \\ \hline
\textbf{NetSquid} &
  \begin{tabular}[c]{@{}l@{}}To simulate all aspects of quantum \\ computing and a quantum network\\ stack.\end{tabular} &
  \begin{tabular}[c]{@{}c@{}}No,\\ but open API\end{tabular} &
  Python &
  \begin{tabular}[c]{@{}c@{}}1G,\\ 2G\end{tabular} &
  \begin{tabular}[c]{@{}l@{}}Pydynaa discrete-event simulation \\ package, with components for \\ protocols and physical models.\end{tabular} &
  \begin{tabular}[c]{@{}l@{}}Errors are encoded as \\ variable hardware parameters.\end{tabular} \\ \hline
\textbf{Quantum Savory} &
  \begin{tabular}[c]{@{}l@{}}To accurately simulate the physical\\ layer dynamics of quantum\\ networks.\end{tabular} &
  Yes &
  Julia &
  \begin{tabular}[c]{@{}c@{}}1G,\\ 2G,\\ 3G\end{tabular} &
  \begin{tabular}[c]{@{}l@{}}Hardware-agnostic, multiformalism\\ framework with support for symbolic\\ algebra and multiple backends.\end{tabular} &
  \begin{tabular}[c]{@{}l@{}}Backend specific error models\\ that can be extended via Kraus\\ operators or Symbolic Algebra\\ System.\end{tabular} \\ \hline
\end{tabular}%
}
\end{table}

Other quantum network simulators have also emerged with more specific objectives and technical approaches.
QuNetSim~\cite{qunetsim} is a quantum network simulator that assumes a packet-based quantum internet and focuses on high-level functionalities, such as routing.
SimQN~\cite{simqn} is a discrete-event-based network simulation platform for quantum network protocols that makes no assumptions about the network architecture.
QKDNetSim~\cite{dervisevic2024large} is a tool for testing novel QKD network management strategies.
ReQuSim~\cite{wallnofer2024faithfully} is designed to be a faithful simulator of first-generation quantum repeaters.
QUreed~\cite{2024qureed} is a physically accurate simulator for quantum network hardware components.
Quditto~\cite{quditto} is an open-source platform that allows researchers to deploy digital twins of QKD network on classical compute hardware. It relies on NetSquid to simulate the quantum behavior of the network.
Also worth mentioning are three simulators from  industry: Aliro Simulator~\cite{aliro}, QNetLab~\cite{qnetlab}, and Q~Sim~\cite{qnetworkdesign}.
Aliro Simulator is a versatile, modular quantum network simulator equipped to model all portions of a quantum network from the smallest optical components to the largest heterogeneous networks.
QNetLab aims at developing quantum network simulations via a no-code approach, that is, a drag-and-drop interface.
Q~Sim is a highly accurate simulator of large-scale quantum networks that comes with a variety of built-in protocols, but it is also extendable and capable of simulating custom protocols.
\section{Software for Operating Quantum Networks} \label{sec:sw-ops}

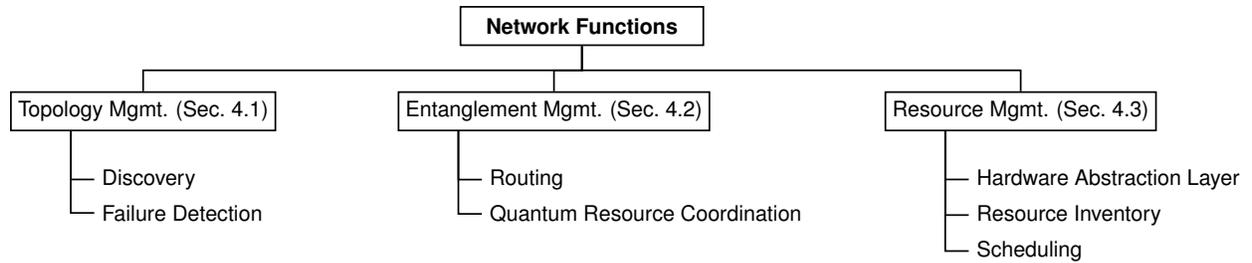
\begin{figure}[htbp]
    \centering
    \resizebox{\columnwidth}{!}{%
    \begin{forest}
        forked edges, folder indent=1cm,
        where={level()<1}{}{folder, grow'=east},
        where={level()>0}{l sep+=1cm}{},
        for tree={
            fork sep=4mm,
            thick, edge=thick,
            font=\sffamily,
            if n children=0{if n=1{yshift=-5mm}{}, for parent={s sep=0mm}}{draw, minimum height=4ex, minimum width=4cm}
        }
        [Network Functions, calign=edge midpoint, s sep=2cm, font=\bfseries\sffamily
            [Topology Mgmt. (Sec.~\ref{sec:topology})
                [Discovery]
                [Failure Detection]
            ]
            [Entanglement Mgmt. (Sec.~\ref{sec:ent_manager})
                [Routing]
                [Quantum Resource Coordination]
            ]
            [Resource Mgmt. (Sec.~\ref{sec:res-mngr})
                [Hardware Abstraction Layer]
                [Resource Inventory]
                [Scheduling]
            ]
        ]
    \end{forest}%
    }
    \caption{Taxonomy of quantum network functions.}
    \label{fig:network_functions_taxonomy}
\end{figure}

As mentioned before, the main service provided by a quantum network is to distribute entanglement between users (and applications) across a wide range of geographical distances.
Following the plane abstraction described in Section~\ref{sec:planes}, a software implementation of the control and service plane is required to operate the network.
A control and service plane  comprises a  network function, a set of operations that collectively achieve a specific objective within the larger network architecture.
Similar to classical networks, a quantum network must perform functions for resource management, synchronization, and topology management in order to deliver entanglement distribution services to users. 
However, these functions must be adapted to accommodate the constraints of quantum information.
Moreover, quantum networks add one non-classical network function: entanglement management.
Figure~\ref{fig:network_functions_taxonomy} proposes a taxonomy of quantum network functions and subfunctions.
It is worth noting that quantum networks for specific purposes (e.g. QKD) may require different architectures and planes. For instance, Lopez et al.~\cite{lopez2025_qkd} proposed an operational model for QKD networks that organizes the network using the following planes: physical, key management, control, and service.

\begin{figure}[htbp]
    \centering
    \includegraphics[scale=0.80]{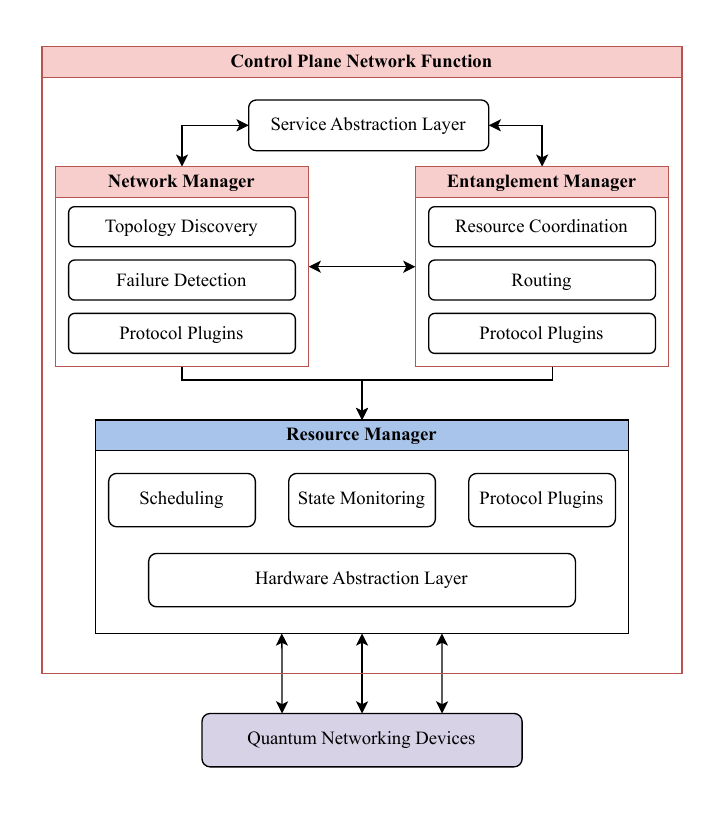}
    \caption{Architecture of network functions and how they interface with the device and user. \textit{Protocol plugins} allow extensibility of network function managers with novel functionality.}
    \label{fig:func_connection}
\end{figure}

As shown in Fig.~\ref{fig:func_connection}, quantum network functions exist as disaggregated components, each having control over their subfunctions. Separate network components should expose interfaces that allow for each network function to base decisions on. For example, the network manager function should expose the topology and channel capabilities to the entanglement manager, where a routing function will then make informed decisions based on the topology. Additionally, each network function  should  expose its control interface that allows extensibility via \textit{protocol plugins}, which are modular components that equip a manager with additional functionality. 
The rest of this section organizes the state of the art in quantum networks following the proposed taxonomy (see Fig.~\ref{fig:network_functions_taxonomy}), with a special focus on software implementation of these functions. 

\subsection{Network Topology Management}~\label{sec:topology}
The  network topology manager  performs discovery and failure detection of links and nodes to build a logical representation of the network topology. In classical networking, many different implementations for topology management are used. Topology discovery/mapping is accomplished by a broad suite of both centralized and decentralized protocols~\cite{kurose_computer_2016}. At the link layer, for example, the ARP protocol maps IP addresses to MAC addresses for local topology discovery. Additionally, the Link Layer Discovery Protocol (LLDP) enables devices to advertise their device information between peers. Network layer protocols such as the Border Gateway Protocol (BGP) and Open Shortest Path First (OSPF) provide decentralized, distributed topology discovery and routing table construction, respectively. The combination of many protocols generates the global network topology distributed across the network. In contrast, in networks following the software-defined networking (SDN) paradigm, centralized network topology protocols are managed by controllers that communicate over an out-of-band channel (e.g., OpenFlow and ForCES) with network devices (e.g., switches, routers, and firewalls) to create a global view of the network~\cite{marin_-depth_2019}.

\subsubsection{Quantum Topology Management Implementations}
Distributed entangled pairs are optical in nature, and thus quantum networks rely on optical channels to distribute their main resource (i.e., entangled photons) while using classical/electro-optical channels for signaling. This difference is the motivation for new topology discovery and management protocols, since traditional methods do not provide means for managing all-optical quantum nodes without destroying quantum states~\cite{main_distributed_2025, schon_quant-net_2024}. 
Theorists have produced a large body of work designing protocols to accomplish specific tasks of the network manager. 
Recently, an increasing number of these protocols have been simulated on quantum network simulators to gain insight into  the network dynamics of such protocols during operation. Table~\ref{tab:network_manager} outlines the implementations discussed in this section.

\paragraph{Hardware-Based Approaches}
In~\cite{main_distributed_2025} the authors  use a configurable all-to-all photonic interconnect to manage topology. This allows the system to dynamically reconfigure itself to an arbitrary topology. In~\cite{schon_quant-net_2024}, the authors implement a non-real-time network server that supplies a global network topology to all nodes on the network. The authors assume that nodes are connected via optical channels, while also having a classical channel to communicate with the control server via remote procedure calls to periodically recalibrate the network, effectively handling optimizations and link failures; however, there are no empirical results for this functionality.

\paragraph{Theoretical Designs}
Several theoretical approaches have been proposed but not yet implemented in simulations or on hardware. In ~\cite{chen_inferring_2023} the authors present a protocol that can build a network topology by using entropic and qubit measurements of an $n$-local quantum network. In~\cite{miguel-ramiro_qping_2025} the authors implement a protocol called QPing that is similar to the ICMP ping protocol of classical networks. The QPing protocol is a diagnostic tool that quantum nodes can use to determine whether entanglement can be generated with sufficient fidelity. However, this is a theoretical design for a protocol and has not been implemented on real hardware or on a simulator. In~\cite{chung_design_2022} the authors design a protocol to discover neighboring nodes connected across a network of all-optical switches.

\paragraph{Simulation Approaches}
In~\cite{mazza_simulation_2024} the authors develop a software implementation of quantum local area networks  by creating two node types: orchestrator and client nodes. The orchestrator nodes distribute multipartite entanglement, and  the client nodes store the state and execute a correction protocol to maintain fidelity. This architecture allows for the simulation of dynamic network topologies that are not limited by physical connectivity constraints. 

\begin{table}[htbp]
    \centering
    \caption{Overview of the protocol implementations for the network manager}
    \begin{tabular}{p{0.25\textwidth}p{0.15\textwidth}p{0.5\textwidth}}
        \toprule
        \textbf{Implementation} & \textbf{Type} & \textbf{Protocol Description} \\
        \midrule
        Main et.\ al.~\cite{main_distributed_2025} & Hardware & All-to-all configurable photonic interconnect \\
        \addlinespace
        Schon et al.~\cite{schon_quant-net_2024} & Hardware & Quantum testbed with control server \\
        \addlinespace
        Chen et al.~\cite{chen_inferring_2023} & Theory & Entropic channel topology discovery\\
        \addlinespace
        Miguel-Ramiro et al.~\cite{miguel-ramiro_qping_2025} & Theory & Dynamic channel charatarization to direct entanglement generation \\
        \addlinespace
        Chung et al.~\cite{chung_design_2022} & Theory & Optical ad hoc topology disovery mechanism \\
        \addlinespace
        Mazza et al.~\cite{mazza_simulation_2024} & Simulation & On-demand topology configuration using virtual links\\
        \bottomrule
    \end{tabular}
    \label{tab:network_manager}
\end{table}

\subsubsection{Future Development and Research Gap}
Although the research community agrees on the necessity for a quantum topology manager, we still lack implementations on real testbeds.
A good starting point is to implement proposed designs of the topology management function on full-stack quantum network simulators. The main gap in the study of the topology management function is the lack of failure scenarios, configuration management, and dynamic topology changes that exist in classical networks. Current topology managers for quantum networks focus on the discovery of channels between node-to-detector and node-to-node~\cite{taherkhani_automatic_2025}  or a topology builder protocol as seen in 
~\cite{chung_design_2022}. In practice, however, a central server shares a global topology with all nodes~\cite{schon_quant-net_2024, pirker_resource-centric_2025}. The topology management function must include protocols that can manage the case of failed nodes and ad hoc configuration. 

As an example, assume a continuous MIM entanglement distribution protocol. In the case that the optical channel is lost or the fidelity of the pairs becomes too low, the topology management function should have the ability to interface with the entanglement management function (see next section) to  perform either rerouting or cancellation of the delivered service. 

\subsection{Entanglement Management}\label{sec:ent_manager}
The  entanglement manager  orchestrates the 
distribution of long-distance, high-fidelity entanglement between users, while minimizing service latency. This network function accepts entanglement requests from users and coordinates the necessary operations to facilitate the distribution of entangled pairs. The entanglement manager takes advantage of routing and scheduling algorithms to manage the link-level entanglement generation, swapping, and purification protocols, assuming the topology manager has established a correct representation of the physical (and logical) topology. 

The fundamental resource orchestrated by the entanglement manager is quantum entanglement.
Since long-distance entanglement is a resource that needs to be requested and quickly consumed, we observe that the entanglement manager could be explained by making analogies to modern 5G network architectures~\cite{dahlman_5g_2018}.
At the physical layer, link-level entanglement shares similarities with channel access in 5G networks. Channel access is granted through orthogonal frequency-division multiple access (OFDMA) resource blocks that divide the spectrum by time and frequency.
At a higher logical level, the functions of the entanglement manager draw parallels to the 5G Session Management Function (SMF). Similar to how the SMF manages the packet data unit (PDU) sessions by establishing, changing, and releasing data paths according to node requests and network conditions, the entanglement manager must coordinate the establishment of paths to fulfill entanglement requests in terms of rate and fidelity. 


\subsubsection{Software Implementations}
Several software implementations of the entanglement manager have been realized as part of a broad quantum network protocol stack. QNodeOS~\cite{delle_donne_operating_2025} implements a full stack operating system for quantum end nodes on a point-to-point nitrogen-vacancy (NV) center (or trapped-ion)  network. The system coordinates entanglement distribution requests using a time division multiple access (TDMA) scheduler that aims to maximize the fidelity of the delivered pairs. In~\cite{dahlberg_link_2019} the authors design and simulate a point-to-point (link layer) entanglement protocol using NetSquid~\cite{coopmans_netsquid_2021}. The Adaptive Continuous entanglement generation Protocol  is a novel protocol that  aims to minimize time to serve entanglement generation requests by continuously generating link-level entanglement while adapting to network conditions. This protocol was implemented and evaluated in the SeQUeNCe quantum network simulator~\cite{zhan_design_2025}. 
Kozlowski et al.~\cite{kozlowski_designing_2020} implemented a virtual circuit protocol to establish links for entanglement generation coupled with entanglement tracking protocols to manage the resource. 
While the quantum wrapper protocol~\cite{ben_yoo_quantum_2024, on_experimental_2024} enables the coexistence of a quantum network with classical networks, the architecture provides broad implementations for entanglement management. The quantum wrapper protocol was designed and implemented on a three-node network, demonstrating the transport of the quantum wrapper datagrams. Table~\ref{tab:ent_manager} outlines the  implementations discussed.

\begin{table}[htbp]
    \centering
    \caption{Overview of software implementations for entanglement management}
    \begin{tabular}{p{0.25\textwidth}p{0.3\textwidth}p{0.35\textwidth}}
        \toprule
        \textbf{Implementation} & \textbf{Technology} & \textbf{Type of Protocol} \\
        \midrule
        Delle Donne et al.~\cite{delle_donne_operating_2025} & NV Center Network & General Quantum Operating System \\
         \addlinespace
        Dahlberg et al.~\cite{dahlberg_link_2019} & Simulation (NetSquid) & Link Layer Protocol\\
         \addlinespace
        Zhan et al.~\cite{zhan_design_2025} & Simulation (SeQUeNCe) & Entanglement Distribution \\
         \addlinespace
        Kozlowski et al.~\cite{kozlowski_designing_2020} & Simulation & Entanglement Routing\\
        \addlinespace
        Yoo et al.~\cite{ben_yoo_quantum_2024, on_experimental_2024} & Optical QNode Network & Network Control and Orchestration\\
        \bottomrule
    \end{tabular}
    \label{tab:ent_manager}
\end{table}

\subsubsection{Routing Protocol Designs}
Routing protocols for quantum networks have been a popular study subject; we refer the reader to~\cite{abane_entanglement_2025} for a comprehensive survey. 
Here we provide a high-level overview of the major types of protocols and their implementations. To accomplish end-to-end entanglement between distant nodes, researchers have proposed three types of routing protocols to perform path computation: proactive, reactive, and hybrid. 
According to~\cite{abane_entanglement_2025}, proactive routing determines the route before any initial entanglement generation is attempted; this creates on-demand resource service. 
Reactive routing continuously produces entanglement across all links in the network. After entanglement has been generated between all links, path computation is performed on the entanglement-based topology to schedule swapping operations that create the final end-to-end links. Hybrid protocols combine features from both reactive and proactive protocols to attempt to optimize the distribution of entanglement. These hybrid approaches have two methods: virtual ~\cite{schoute_shortcuts_2016} and opportunistic. Each of these approaches utilizes a controller to respond to and manage entanglement requests between users.
Finally, on a long-term vision of the quantum internet architecture, Caleffi and Cacciapuoti~\cite{caleffi2025quantum-native} proposed a quantum-native routing protocol that leverages a novel quantum addressing scheme that encodes device addresses as quantum states on the header of an entanglement packet.

\subsection{Resource Management and Scheduling} \label{sec:res-mngr}
In physical quantum systems, resource managers need to provide protocols with access to quantum hardware resources. 
Classical computing systems typically achieve this hardware abstraction through hardware abstraction layers (HALs), which provide standardized interfaces to diverse hardware components. 
In the SeQUeNCe quantum network simulator, the role of the resource manager is to manage local resources by implementing state tracking and control flow to direct the execution of the protocol~\cite{wu_sequence_2021}.
On the QUANT-NET testbed, Yu et al.~\cite{yu_extensible_2025} implemented a quantum network control plane (QNCP) that implements a hardware abstraction layer, a quantum resource management function, and an orchestration layer using centralized control architecture. 
In QNCP, the centralized controller is responsible for maintaining knowledge of the resources, states, and configurations of the network components, 
while the HAL is implemented within the distributed agent components and provides control logic for accessing hardware devices. 
Alternatively, in~\cite{pirker_resource-centric_2025} the authors propose a resource-centric, task-based quantum network control framework that implements a network resource manager, which is designed as the central coordination point for deriving distributed workflows from application objectives.

Although proposals for a quantum internet foresee asynchronous operations, the current consensus for near-term quantum network implementations identifies scheduling as a key network function.
In~\cite{yu_two-level_2025} the authors propose a synchronous time-slot scheduling scheme, in which a central controller allocates and schedules device-specific quantum network operations to time slots (similar to TDMA). 
Additionally, a local scheduler runs at the node level and  is responsible for scheduling real-time functions such as periodic calibrations. 
This scheduling framework was evaluated on the QUANT-NET~\cite{schon_quant-net_2024} testbed.
Similarly, Beauchamp et al.~\cite{beauchamp_extended_2025} develop a network architecture that enables the scheduling of network operations by applications to facilitate entanglement distribution. 
The architecture considers the \textit{generate-when-requested} entanglement generation scheme, where entanglement is distributed to nodes on demand, instead of continuously. 
The end nodes contain a quantum network agent that interfaces with a centralized SDN controller. 
This controller computes a network scheduling flow, which will in turn gives the network schedule back to the end nodes. 
The protocols for this architecture were simulated on a six-node star topology using an ad hoc simulator. Table~\ref{tab:resource_manager} provides a summary of the discussed resource management functions and implementations.

\begin{table}[htbp]
    \centering
    \caption{Overview of resource management implementations}
    \begin{tabular}{p{0.2\textwidth}p{0.7\textwidth}}
        \toprule
        Implementation & Method \\
        \midrule
        Yu et al.~\cite{yu_extensible_2025} & Control plane resource management on a physical testbed. \\
        \addlinespace
        Wu et al.~\cite{wu_sequence_2021} & Local state tracking and protocol control flow in simulation. \\
        \addlinespace
        Pirker et al.~\cite{pirker_resource-centric_2025}& Global resource manager -- composes network objective  by assigning tasks to quantum and classical resources. \\
        \addlinespace
        Yu et al.~\cite{yu_two-level_2025}& Two-level framework for real-time (node-level) and non-real-time (network-level) task scheduling. \\
        \addlinespace
        Beauchamp et al.~\cite{beauchamp_extended_2025} & End nodes -- submit requests  to a centralized controller in charge of computing a global network schedule. \\
        \bottomrule
    \end{tabular}
    
    \label{tab:resource_manager}
\end{table}
\section{Discussion} \label{sec:discussion}
In this paper we presented a review of software for quantum networks following a plane abstraction.
Our main contribution is to organize the state of the art in terms of the functions that a quantum network must execute to reliably provide entanglement distribution services.
Among the major functions (i.e., topology management, entanglement management, and resource management), our review shows that entanglement management has received the most attention, with numerous theoretical and simulation studies proposing protocols. 
However, software implementations of these proposals on real quantum network testbeds remain sparse.
While quantum hardware components are still immature, implementing designs on realistic simulators remains a critical step toward bridging this gap.

Looking ahead, several specific areas merit further attention. First, topology management protocols must evolve beyond discovery toward handling failures, dynamic reconfiguration, and ad hoc connectivity. Second, resource management and scheduling frameworks should explicitly account for quantum-specific constraints such as memory decoherence, and probabilistic entanglement generation. Third, standardized interfaces and software-defined orchestration tools are needed to enable interoperability across heterogeneous quantum devices. Moreover, there is an opportunity to extend today’s control/service plane with a management plane that incorporates configuration management, software updates, and health monitoring for scalability.

Encouragingly, the quantum internet research group  of the  Internet Engineering Task Force is already working toward common interfaces and data formats~\cite{lopez-qirg-qi-multiplane-arch-04}. Continued collaboration of software developers, experimentalists, and standards bodies will be essential for advancing scalable software architectures that can support hybrid, large-scale quantum networks.

\medskip
\textbf{Acknowledgments} \par
This material is based upon work supported by the U.S.~Department of Energy, Office of Science, Advanced Scientific Computing Research (ASCR) program under contract number DE-AC02-06CH11357 as part of the InterQnet quantum networking project.

\newpage
\textbf{Government License} \par
The submitted manuscript has been created by UChicago Argonne, LLC, Operator of Argonne National Laboratory (``Argonne''). Argonne, a U.S. Department of Energy Office of Science laboratory, is operated under Contract No. DE-AC02-06CH11357. The U.S. Government retains for itself, and others acting on its behalf, a paid-up nonexclusive, irrevocable worldwide license in said article to reproduce, prepare derivative works, distribute copies to the public, and perform publicly and display publicly, by or on behalf of the Government.  The Department of Energy will provide public access to these results of federally sponsored research in accordance with the DOE Public Access Plan. \href{http://energy.gov/downloads/doe-public-access-plan}{http://energy.gov/downloads/doe-public-access-plan.}

\medskip

\bibliographystyle{IEEEtran}
\bibliography{bibliography}

\end{document}